\RequirePackage[displaymath]{lineno}
\documentclass[aps,prc,twocolumn,showpacs,superscriptaddress,floatfix,nofootinbib]{revtex4-2}
\usepackage{newtxtext,newtxmath,booktabs,siunitx}
\usepackage{dcolumn}
\usepackage{bm}
\usepackage{float}
\usepackage{ulem}

\usepackage{graphicx}
\usepackage{booktabs}
\usepackage{tabularx}
\usepackage{amsmath}
\usepackage{xcolor}
\usepackage{multirow}
\usepackage[colorlinks,citecolor=blue,urlcolor=blue,linkcolor=blue]{hyperref}

\newcommand{\ac}{{\rm ac}_{2}\{3\}}
\newcommand{\cosDPhi}{\cos4(\Phi_4-\Phi_2)}

\newcommand{\vtt}{v_{2}\{2\}}

\newcommand{\vft}{v_{4}\{2\}}
\newcommand{\snn}{\sqrt{s_{\rm NN}}}

\newcommand{\AuAu}{$^{197}$Au+$^{197}$Au}
\newcommand{\UU}{$^{238}$U+$^{238}$U}

\begin{document}

\title{Investigation of Nonlinear Collective Dynamics in Relativistic Heavy-Ion Collisions Using A Multi-Phase Transport Model}

\author{Zhi-Jie Yang}
\affiliation{College of Science, Wuhan University of Science and Technology, Wuhan, Hubei 430065, China}
\affiliation{School of Science, Huzhou Normal University, Huzhou, Zhejiang 313000, China}

\author{Hao-jie Xu}
\email{haojiexu@zjhu.edu.cn}
\affiliation{School of Science, Huzhou Normal University, Huzhou, Zhejiang 313000, China}
\affiliation{Strong-Coupling Physics International Research Laboratory (SPiRL), Huzhou Normal University, Huzhou, Zhejiang 313000, China.}

\author{Jie Zhao}
\email{jie_zhao@fudan.edu.cn}
\affiliation{Key Laboratory of Nuclear Physics and Ion-beam Application (MOE), Institute of Modern Physics, Fudan University, Shanghai 200433, China}
\affiliation{Shanghai Research Center for Theoretical Nuclear Physics, NSFC and Fudan University, Shanghai 200438, China.}

\author{Hanlin Li}
\email{lihl@wust.edu.cn}
\affiliation{College of Science, Wuhan University of Science and Technology, Wuhan, Hubei 430065, China}

\begin{abstract}
The nonlinear response coefficient, $\chi_{4,22}$, is a crucial observable for probing the dynamical properties of the quark-gluon plasma (QGP). Although traditionally interpreted as a signature of medium response, recent studies suggest that $\chi_{4,22}$ also encodes information about the intrinsic initial-state configuration of the colliding nuclei. In this study, we use the A Multi-Phase Transport (AMPT) model to investigate the microscopic origin and stage-by-stage development of $\chi_{4,22}$ in $^{238}$U+$^{238}$U and $^{197}$Au+$^{197}$Au collisions at $\snn = 200$ GeV. By tracking flow observables through the partonic cascade, quark coalescence, and hadronic rescattering phases, we map the conversion of initial geometric eccentricities into final-state momentum anisotropies. Our results demonstrate that the absolute magnitude of $\chi_{4,22}$ increases continuously during collective expansion, confirming its nature as a dynamically generated medium response. In contrast, the relative ratio of this coefficient between the U+U and Au+Au systems, $R(\chi_{4,22})$, exhibits approximate stage independence over a broad centrality range, as quantified by a constant-fit test across the three AMPT evolution stages. This indicates that the ratio reduces common medium-response effects, such as overall amplification efficiency and viscous attenuation, and therefore retains stronger sensitivity to the relative initial-state geometry. These findings provide theoretical support for using nonlinear flow ratios to constrain higher-order nuclear structure, such as hexadecapole deformation, while also clarifying the residual stage dependence and nonflow limitations relevant to precision extractions.
\end{abstract}

\maketitle

\section{Introduction}

Relativistic heavy-ion collisions recreate a state of matter in which quarks and gluons are deconfined over nuclear length scales---the quark-gluon plasma (QGP)---analogous to the conditions present in the early universe microseconds after the Big Bang~\cite{Adams:2005dq,Adcox:2004mh,Shuryak:1980tp}. Experimental data from these collisions indicate strong collective expansion, suggesting that the QGP behaves as a strongly coupled quantum chromodynamics fluid~\cite{Kolb:2003dz,Heinz:2013th, Gale:2013da,Jeon:2015dfa,Song:2017wtw}. A key challenge in heavy-ion experiments is that observables are measured only in the final state, whereas the QGP exists only during the early stages of the collision---typically beginning at a timescale of $\sim 0.5$ fm/$c$. Consequently, our understanding of the QGP relies entirely on mapping the initial state to the final state through the dynamical evolution of the collision system.

Although this mapping is inherently complex, dynamical models based on hydrodynamics and transport theory have established a successful relationship between initial spatial eccentricities ($\epsilon_{n}$) and final anisotropic flow observables ($v_{n}$)~\cite{Ollitrault:1992bk,Kolb:2000fha}. Specifically, a linear relationship, $v_n \propto \epsilon_n$, holds well for $n=2, 3$~\cite{Qiu:2011iv,Wei:2018xpm}. Such a mapping simplifies the analysis of flow observables by enabling the use of computationally efficient initial-state models. This is particularly important for recent studies of nuclear structure in relativistic heavy-ion collisions, where large statistical ensembles of events must be generated to extract subtle sensitivities~\cite{Giacalone:2021udy,Xu:2024bdh}.

For higher-order flow harmonics such as $v_4$ and $v_5$, this linear approximation is no longer sufficient, and nonlinear response mechanisms become important~\cite{Yan:2015jma}. For example, $v_4$ can receive a substantial contribution from the lower-order harmonic $v_2$. In early studies, the nonlinear response coefficient, $\chi_{4,22}$, was regarded as arising solely from the medium response and was assumed to be independent of the initial-state configuration. However, recent studies indicate that $\chi_{4,22}$ is sensitive to the hexadecapole deformation ($\beta_4$) of the colliding nuclei~\cite{Xu:2024bdh,Ryssens:2023fkv}. Preliminary results from the STAR collaboration point to a sizable $\beta_4$ deformation, making a precise investigation of these nonlinear response behaviors timely and essential~\cite{Zhao:2026xyz}. Owing to the nature of the medium response, studying these nonlinear dynamics in principle requires simulating the full medium evolution, which demands substantial computational resources.

To mitigate these theoretical uncertainties, nuclear structure studies in heavy-ion collisions often rely on ratio observables between two reference systems, such as relativistic isobar collisions or semi-isobar pairs like \UU\ and \AuAu~\cite{Xu:2017zcn,Li:2019kkh,Giacalone:2021udy,JIa:2021wbq}. 
In such comparative measurements, uncertainties arising from the system evolution---such as the damping effects of shear and bulk viscosities in hydrodynamic simulations---are expected to be reduced. Nevertheless, for a complex observable such as the nonlinear response coefficient $\chi_{4,22}$, it remains critical to understand its stage-by-stage development during the medium evolution.

Microscopic transport models provide a complementary and powerful framework for addressing this issue. In the A Multi-Phase Transport (AMPT) model~\cite{Lin:2004en}, the collision system evolves through several distinct stages: initial partonic scatterings described by Zhang's Parton Cascade (ZPC)~\cite{Zhang:1997ej}, hadronization via quark coalescence, and subsequent hadronic rescattering. This multi-stage structure enables a direct, time-resolved investigation of how nonlinear response coefficients such as $\chi_{4,22}$ are generated and modified throughout the entire system evolution. Therefore, the primary goal of this work is to present a systematic study of $\chi_{4,22}$ and related observables within the AMPT framework. By isolating the specific contributions from the partonic and hadronic stages, this study provides insight into the microscopic origin of nonlinear collective behavior in relativistic heavy-ion collisions.

The remainder of this paper is organized as follows. In Sec.~II, we introduce the methodology and observables used in this study, detailing the initialization of the deformed Woods-Saxon geometry, the mathematical formulation of the nonlinear response coefficient $\chi_{4,22}$, and the specific configurations of the AMPT model used to track the evolution of the collision system. In Sec.~III, we present our results and discussion, examining the stage-by-stage development of the flow observables and testing the extent to which the U+U to Au+Au ratio method reduces common medium-evolution effects while preserving sensitivity to the initial hexadecapole deformation. Finally, a summary of our findings and their implications is provided in Sec.~IV.

\section{Methodology and Observables}

\subsection{Initial Collision Geometry and Nuclear Deformation}

The spatial anisotropy present at the onset of a heavy-ion collision is governed by three main factors: the macroscopic collision geometry determined by the impact parameter, the intrinsic shape deformations of the colliding nuclei, and event-by-event quantum fluctuations in the positions of the constituent nucleons. To model the initial density profile, the nucleon distribution is typically parameterized using a deformed Woods-Saxon density function~\cite{Woods:1954zz,Filip:2009zz}:
\begin{equation}
\rho(r,\theta,\phi) = \frac{\rho_0}{1 + \exp\left[\frac{r - R(\theta,\phi)}{a}\right]}\,,
\end{equation}
where $\rho_0$ represents the nuclear saturation density, and $a$ denotes the surface diffuseness. The structural deformations of the nucleus are encoded in the radius parameter $R(\theta,\phi)$, which is expanded in spherical harmonics. Under the assumption of axial symmetry, this expansion retains only the $m=0$ terms:
\begin{equation}
R(\theta,\phi) = R_0 \left[ 1 + \sum_{\ell} \beta_{\ell} Y_{\ell 0}(\theta,\phi) \right]\,.
\end{equation}
The deformation parameters, $\beta_{\ell}$, determine geometric variations in the nuclear surface across different multipole orders. Specifically, $\beta_2$ describes quadrupole deformation (corresponding to prolate or oblate shapes), $\beta_3$ represents reflection-asymmetric octupole (pear-like) shapes, and $\beta_4$ accounts for hexadecapole deformations that modify the surface curvature beyond the leading quadrupole term. These intrinsic geometric features are directly imprinted on the initial energy density of the collision zone and act as the primary geometric drivers of subsequent collective-flow observables in the most central collisions. 

In this work, we simulate \UU\ and \AuAu\ collisions at a center-of-mass energy of $\snn = 200$ GeV. To isolate the effects of the geometry of the uranium nucleus, its initial density profile is sampled with specific quadrupole and hexadecapole deformation parameters set to $\beta_2 = 0.286$ and $\beta_4 = 0.100$~\cite{Raman:2001nnq,Ryssens:2023fkv}. For the Au nucleus, we use the nonzero deformation parameters $\beta_{2}=-0.131$ and $\beta_{4}=0.031$, following the nuclear-structure inputs adopted in previous Glauber/AMPT studies~\cite{Lin:2004en,Miller:2007ri,Loizides:2017ack}. Thus, Au+Au serves as an approximately spherical reference system only in comparison with the strongly deformed U+U system. Collision centrality is determined from the final-state charged-particle multiplicity at midrapidity, following standard experimental procedures. The same final-state centrality classification is then applied to the earlier partonic and coalescence-stage samples, ensuring that the same centrality-selected event ensembles are compared throughout the evolution.

\subsection{Flow Observables and Nonlinear Response}

In relativistic heavy-ion collisions, the azimuthal momentum anisotropy of emitted particles is quantified by the Fourier flow harmonics, $v_n$~\cite{Poskanzer:1998yz}. For higher-order harmonics ($n \geq 4$), such as the hexadecapole flow coefficient $v_4$, the observed fourth-order flow vector is commonly decomposed into linear and nonlinear components:
\begin{equation}
V_4 = V_{4L} + V_4^{\rm (NL)} = V_{4L} + \chi_{4,22} V_2^2\,.
\end{equation}
Here, $V_{4L}$ denotes the component that responds directly to the fourth-order initial eccentricity, whereas $V_4^{\rm (NL)}$ is the nonlinear mode-coupled contribution generated from the lower-order elliptic flow vector $V_2$ during the medium evolution. The strength of this coupling is characterized by the nonlinear response coefficient $\chi_{4,22}$. Because the simple linear relation $v_n \propto \epsilon_n$ breaks down for $n \geq 4$, a dynamical description of the collision medium is required to understand how the linear and nonlinear components are generated and modified.

To isolate these nonlinear contributions, we examine observables calculated with respect to the second-order event plane ($\Phi_2$) rather than the same-order plane ($\Phi_4$). The flow-harmonic correlation between $v_2$ and $v_4$ is effectively captured by the three-particle asymmetric cumulant $\ac$, defined as~\cite{Bilandzic:2010jr}:
\begin{equation}
\ac \equiv \langle\langle 3 \rangle\rangle_{2,2,-4} = \langle v_2^4 \rangle^{1/2} v_4\{\Phi_2\}\,.
\end{equation}
Here,
\begin{equation}
\langle v_2^4 \rangle
\equiv
\langle\!\langle 4 \rangle\!\rangle_{2,2,-2,-2}
=
2v_2\{2\}^4-v_2\{4\}^4
\end{equation}
denotes the four-particle cumulant. In the absence of non-flow effects, $\ac$ can be approximated analytically as a flow angular correlation:
\begin{equation}
\ac \approx \langle v_2^2 v_4 \cosDPhi \rangle\,.
\end{equation}
This correlation isolates the projection of $v_4$ onto the $\Phi_2$ event plane. The nonlinear response coefficient is then obtained by normalizing the asymmetric cumulant:
\begin{equation}
\chi_{4,22} \equiv \frac{v_4\{\Phi_2\}}{\langle v_2^4 \rangle^{1/2}} = \frac{\ac}{\langle v_2^4 \rangle}\,.
\label{eq:chi}
\end{equation}
While lower-order cumulants and $\ac$ are sensitive to both quadrupole ($\beta_2$) and octupole ($\beta_3$) deformations, previous studies have shown that $\chi_{4,22}$ is comparatively more sensitive to hexadecapole deformation ($\beta_4$)~\cite{Xu:2024bdh}. In the present work, we do not perform an independent scan over $\beta_4$. Instead, we use AMPT to test whether the corresponding U+U to Au+Au ratio remains sufficiently stable across different evolution stages to support its use as a nuclear-structure-sensitive observable.

To mitigate theoretical uncertainties arising from the dynamical evolution of the collision medium, we construct relative observables between two similar collision systems. In this study, we use Au+Au collisions as a baseline for comparison with U+U collisions:
\begin{equation}
R(X) = \frac{X_{\rm UU}}{X_{\rm AuAu}}\,,
\end{equation}
where $X$ represents a given observable (e.g., $\vtt^2$, $\ac$, or $\chi_{4,22}$). Because U+U and Au+Au are simulated with the same transport setup and analyzed using the same procedure, common medium-response effects are expected to be reduced in this ratio. This cancellation is approximate rather than exact, because residual differences in system size, deformation, nonflow contributions, and centrality dependence can still contribute.

\subsection{A Multi-Phase Transport (AMPT) Model}

To trace the dynamical origins of the nonlinear response coefficients and understand how initial-state spatial eccentricities are converted into final-state momentum anisotropies, we employ the AMPT model~\cite{Lin:2004en}. AMPT is a comprehensive hybrid Monte Carlo framework that simulates the full evolution of relativistic heavy-ion collisions.

Two distinct hadronization mechanisms are available within the AMPT framework: the default mode, which relies on Lund string fragmentation, and the string-melting mode, in which all excited strings are fully converted into their constituent quarks and antiquarks. In this analysis, we use the string-melting version because it is more suitable for describing the partonic dynamics and strong collective flow observed at intermediate and high collision energies~\cite{Bzdak:2014dia,He:2017tla,Zhang:2018ucx,Lin:2021mdn}. The modular structure of AMPT allows us to extract snapshots of the collision system at distinct stages of its evolution, thereby disentangling the contributions of the partonic and hadronic phases. We focus on the system's evolution across the following key stages. We generated a large sample of $1\times10^7$ minimum-bias events for each collision system.

The observables are calculated with the same pseudorapidity acceptance, $|\eta|<2.0$, for the available degrees of freedom at each stage: partons at the partonic stage and charged hadrons at the coalescence and final stages. This wider pseudorapidity window, compared with the commonly used midrapidity cut of $|\eta|<1.0$, is adopted to improve the statistical precision of the multiparticle correlation observables. The same selection is applied to U+U and Au+Au collisions and to the corresponding evolution stages. The statistical uncertainties shown by the shaded bands are estimated by dividing the full event sample into independent subsamples and calculating the standard error of the resulting observable distribution. For ratio observables, the ratio is constructed in each subsample before the statistical uncertainty is evaluated.

\subsubsection{Partonic Phase}
The initial spatial configuration is generated by the Heavy Ion Jet Interaction Generator (HIJING)~\cite{Wang:1991hta}, which samples the deformed Woods-Saxon profiles described above. Soft particle production is governed by the Lund symmetric fragmentation function, $f(z) \propto z^{-1}(1-z)^a \exp(-b m_\perp^2/z)$, with the parameters set to $a = 0.55$ and $b = 0.15$ GeV$^{-2}$. In the string-melting scenario, these excited strings are converted into partons. The subsequent interactions among the free partons are simulated by Zhang's Parton Cascade (ZPC)~\cite{Zhang:1997ej}, which accounts for two-body elastic scatterings. For our simulations at $\snn = 200$ GeV, the strong coupling constant is set to $\alpha_s = 0.33$, and the Debye screening mass to $\mu_D = 3.2032$ fm$^{-1}$, yielding a total partonic scattering cross section of $\sigma = 1.5$ mb. Extracting the observables immediately after the ZPC stage isolates the pure response of the partonic medium.

\subsubsection{Hadronization} 
Once the partonic medium expands and cools, quarks and antiquarks recombine into hadrons via a spatial quark coalescence mechanism. By calculating the flow observables immediately after this coalescence phase, we can investigate how the kinematic recombination of partons into bound hadronic states modifies the previously established momentum anisotropies. To isolate the coalescence-stage observables and effectively remove the influence of the subsequent hadronic cascade, we cap the evolution time at $t_{\rm max} = 0.6$ fm/$c$ in this intermediate scenario. This value is used as a technical AMPT setting, following the recommendation in the AMPT package documentation/readme, to efficiently suppress hadronic rescattering; it should not be interpreted as a physical hadronization timescale.

\subsubsection{Full Evolution Including Hadronic Rescattering}
The final stage of the collision involves the evolution of the newly formed hadronic matter, which is described by the A Relativistic Transport (ART) model~\cite{Li:1995pra} with its default energy-dependent hadronic interaction cross-sections. This phase accounts for elastic and inelastic hadronic scatterings, resonance decays, and baryon-antibaryon annihilations. The system evolves continuously until it reaches kinetic freeze-out, with the evolution time typically limited to $t_{\rm max} = 30$ fm/$c$. Comparing the observables extracted after this full evolution with those from earlier stages provides a measure of how final-state hadronic rescattering modifies the nonlinear response.

\begin{figure*}[!thb]
    \centering
    \includegraphics[width=0.45\textwidth]{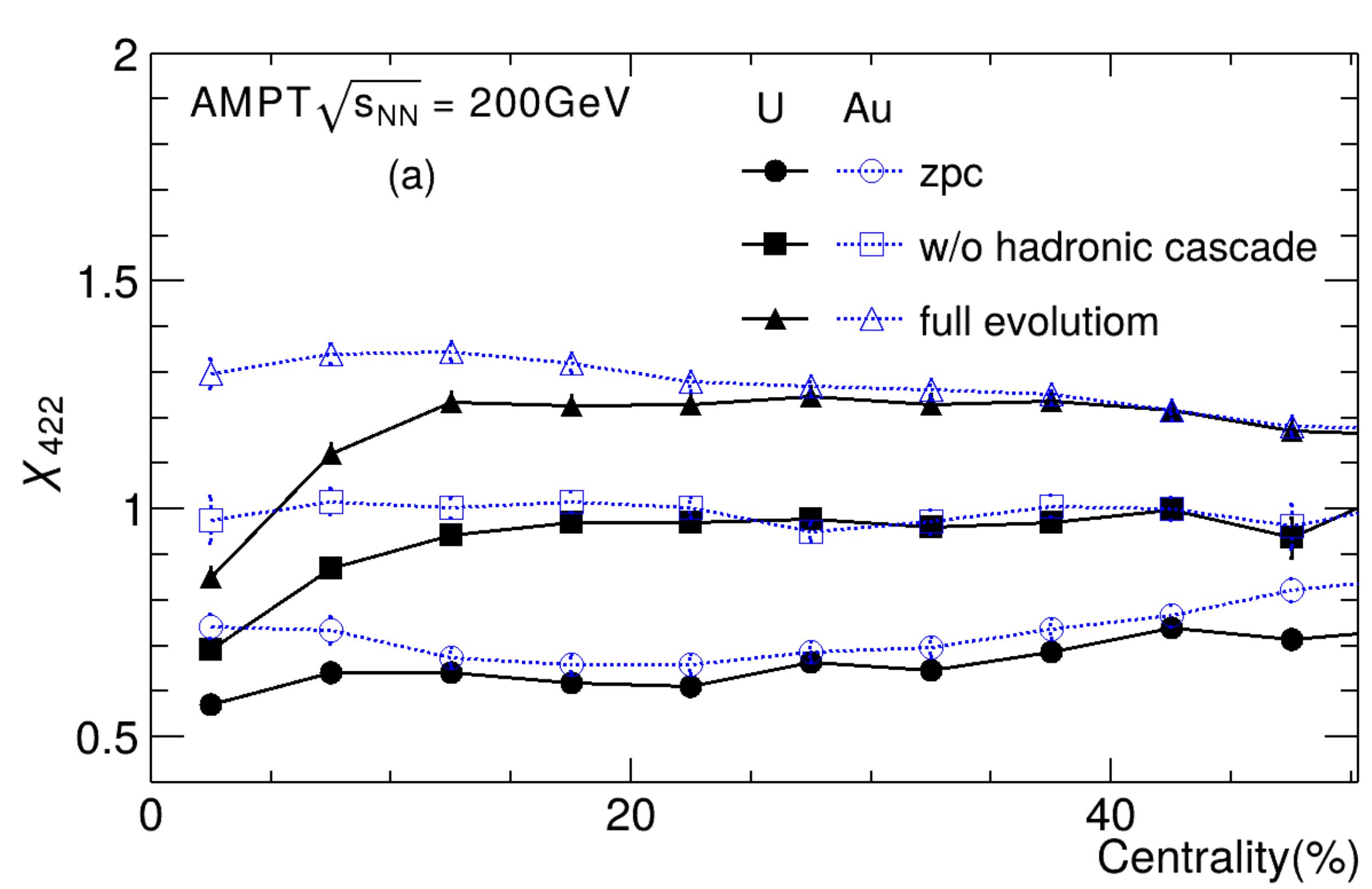}  
    \includegraphics[width=0.45\textwidth]{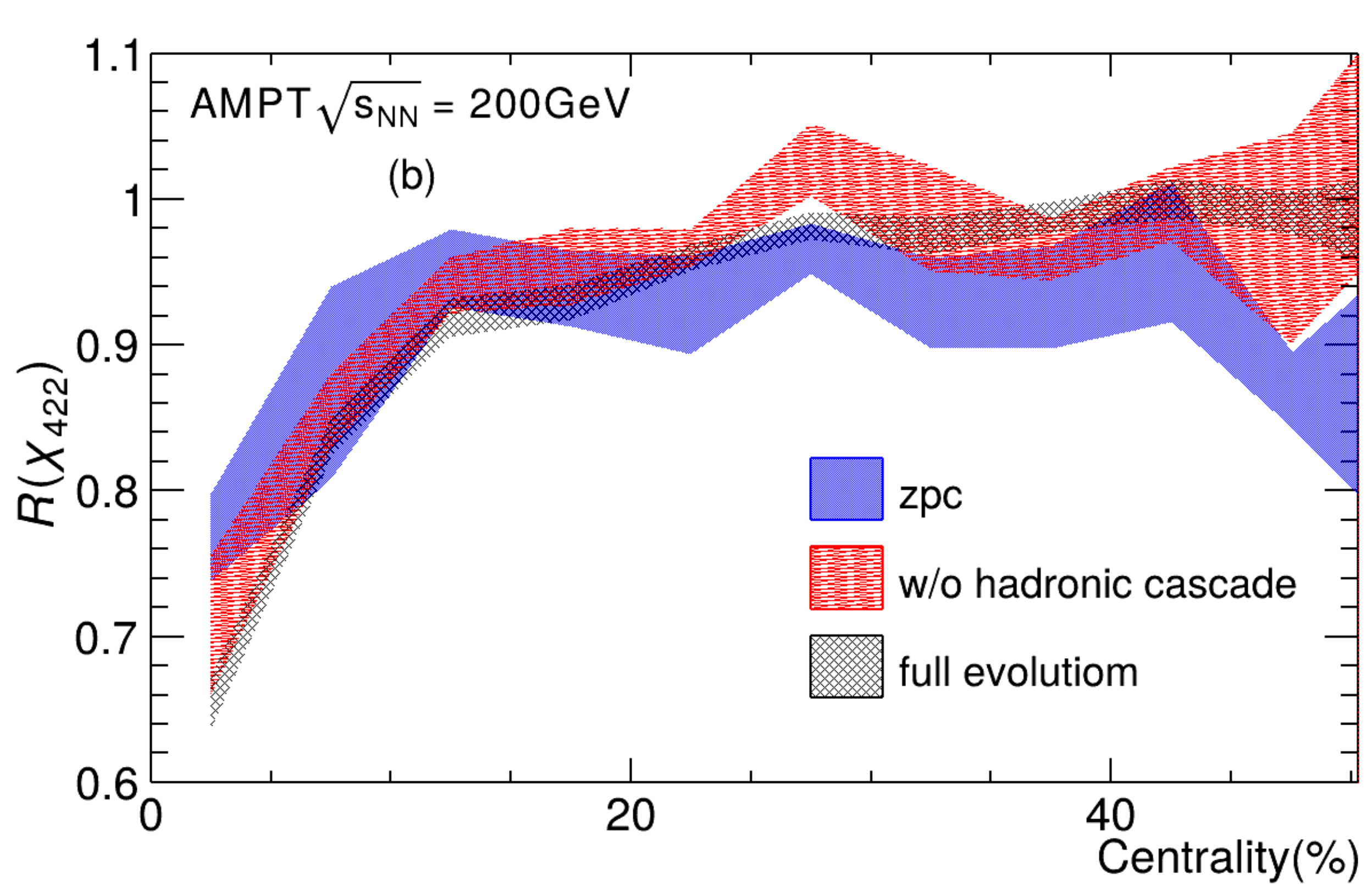}  
    \caption{Centrality dependence of the nonlinear response coefficient $\chi_{4,22}$ extracted from AMPT simulations of U+U and Au+Au collisions at $\snn = 200$ GeV. Panel (a) shows the absolute magnitude of $\chi_{4,22}$ for each collision system, whereas panel (b) presents the corresponding relative ratio between the two systems. Results are shown for three distinct stages of the system evolution: immediately after the partonic cascade (zpc), after hadronization via quark coalescence (w/o hadronic cascade), and after the full evolution, including final-state hadronic rescattering (full evolution). Shaded bands represent statistical uncertainties estimated from independent subsamples.}
    \label{fig:chi422}
\end{figure*}

\begin{figure*}[!hbt]
    \centering
    \includegraphics[width=0.45\textwidth]{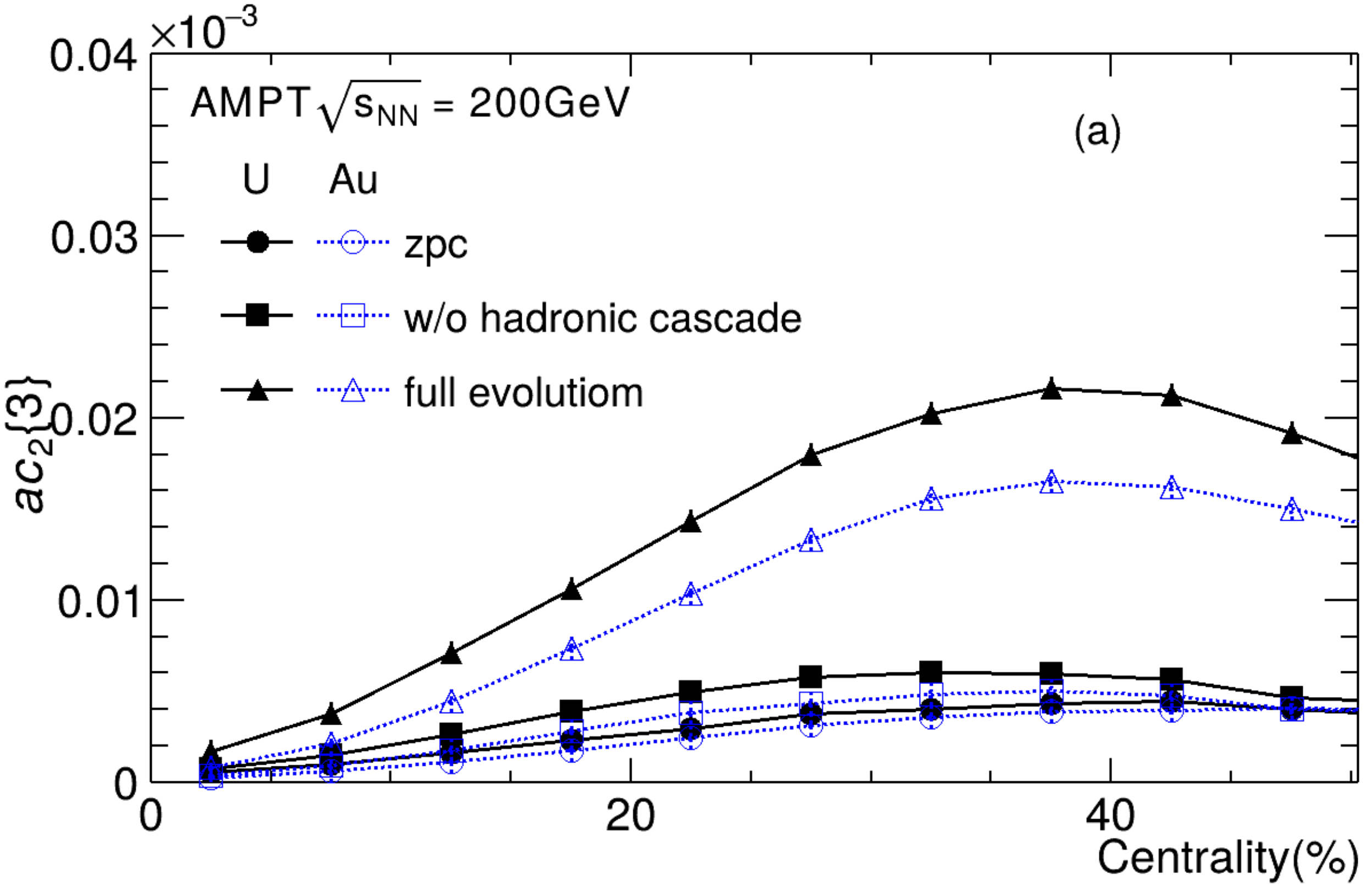} 
    \includegraphics[width=0.45\textwidth]{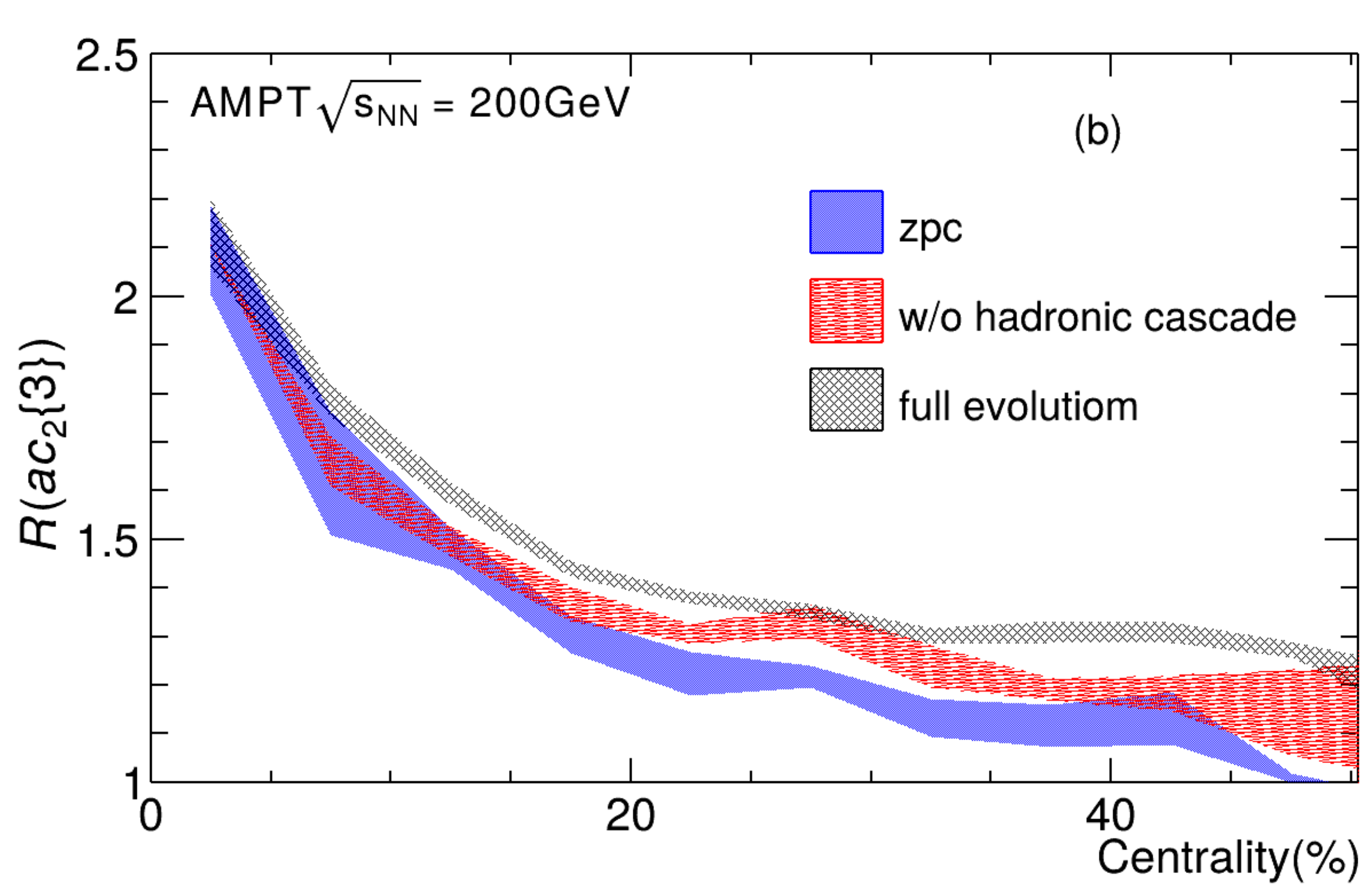}  \\
    \includegraphics[width=0.45\textwidth]{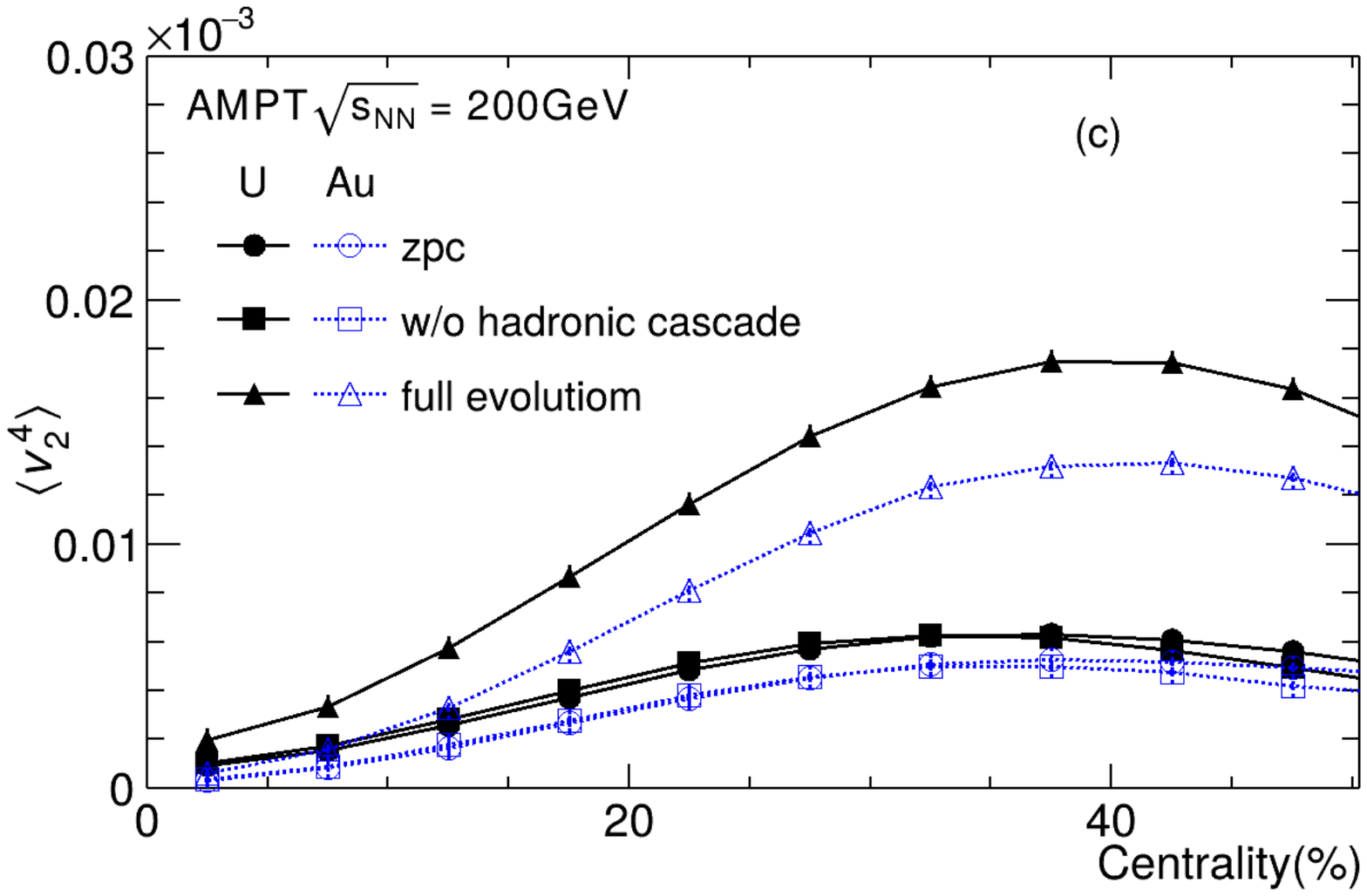} 
    \includegraphics[width=0.45\textwidth]{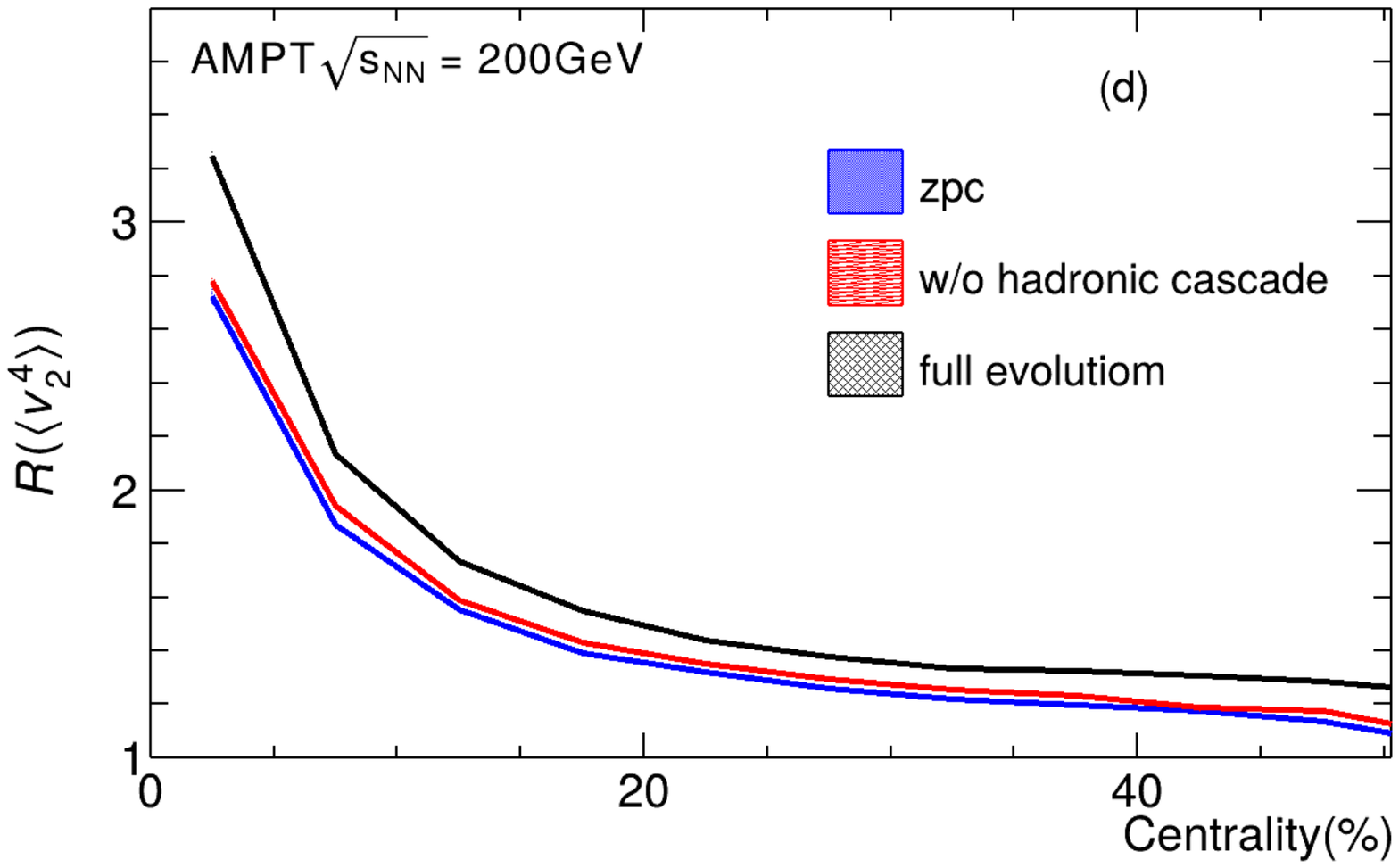}  
    \caption{Same as Fig.~\ref{fig:chi422}, but for the three-particle asymmetric cumulant $\ac$ (a and b) and the four-particle cumulant $\langle v_{2}^{4} \rangle$ (c and d). The shaded bands indicate statistical uncertainties.}
    \label{fig:acandv24}
\end{figure*}

\begin{figure*}[!hbt]
    \centering
    \includegraphics[width=0.45\textwidth]{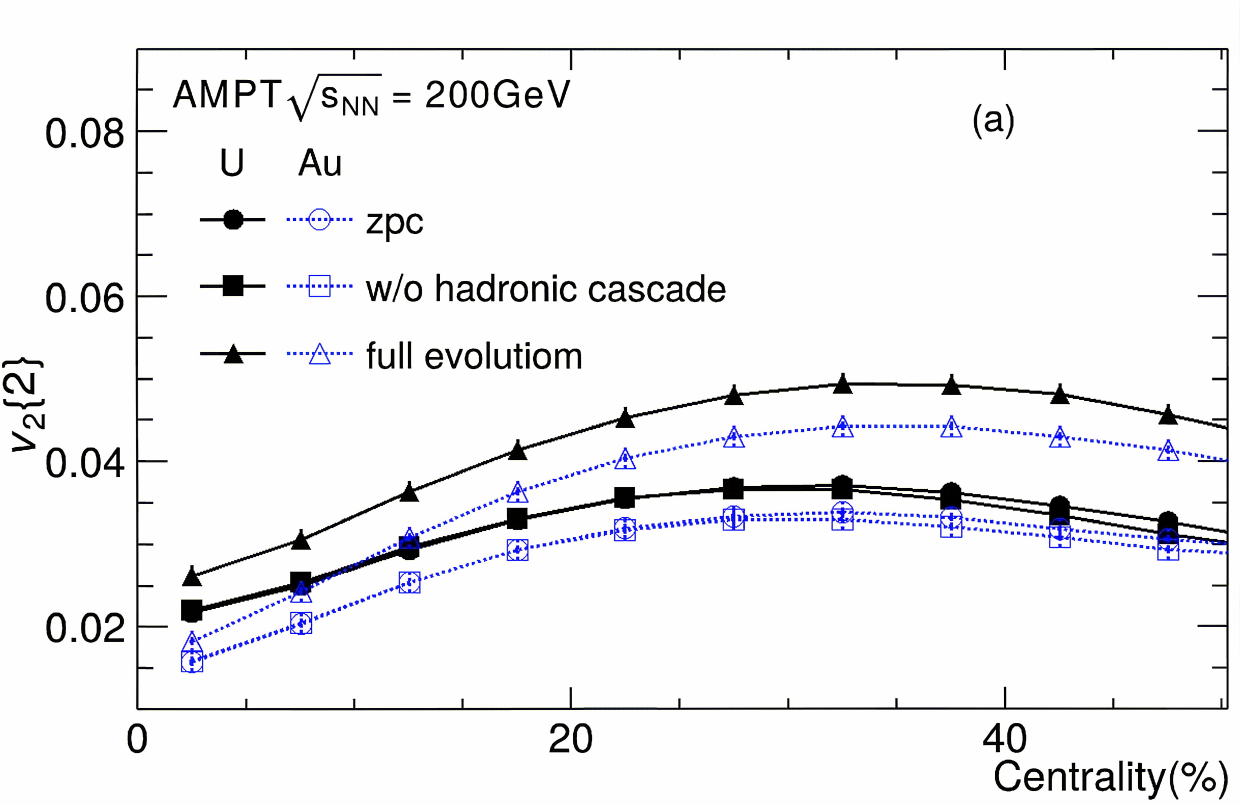}  
    \includegraphics[width=0.45\textwidth]{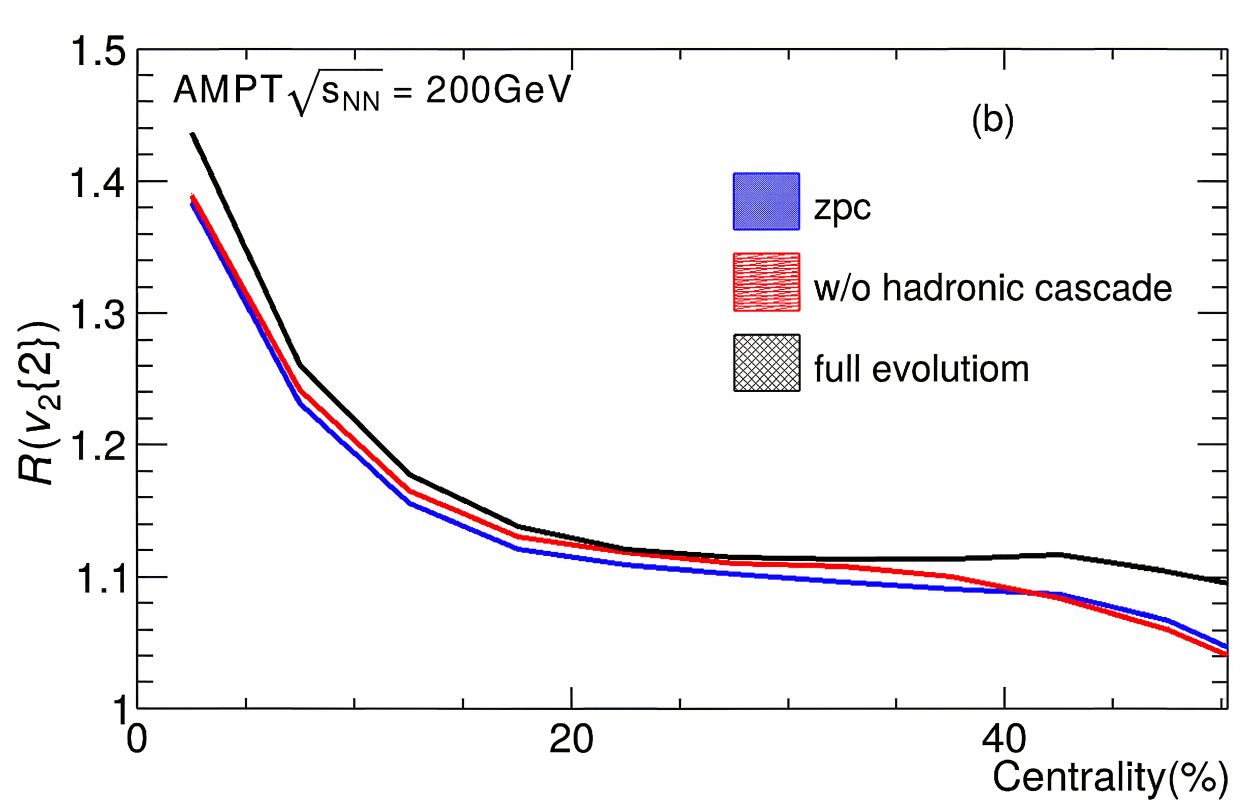}  \\
    \includegraphics[width=0.45\textwidth]{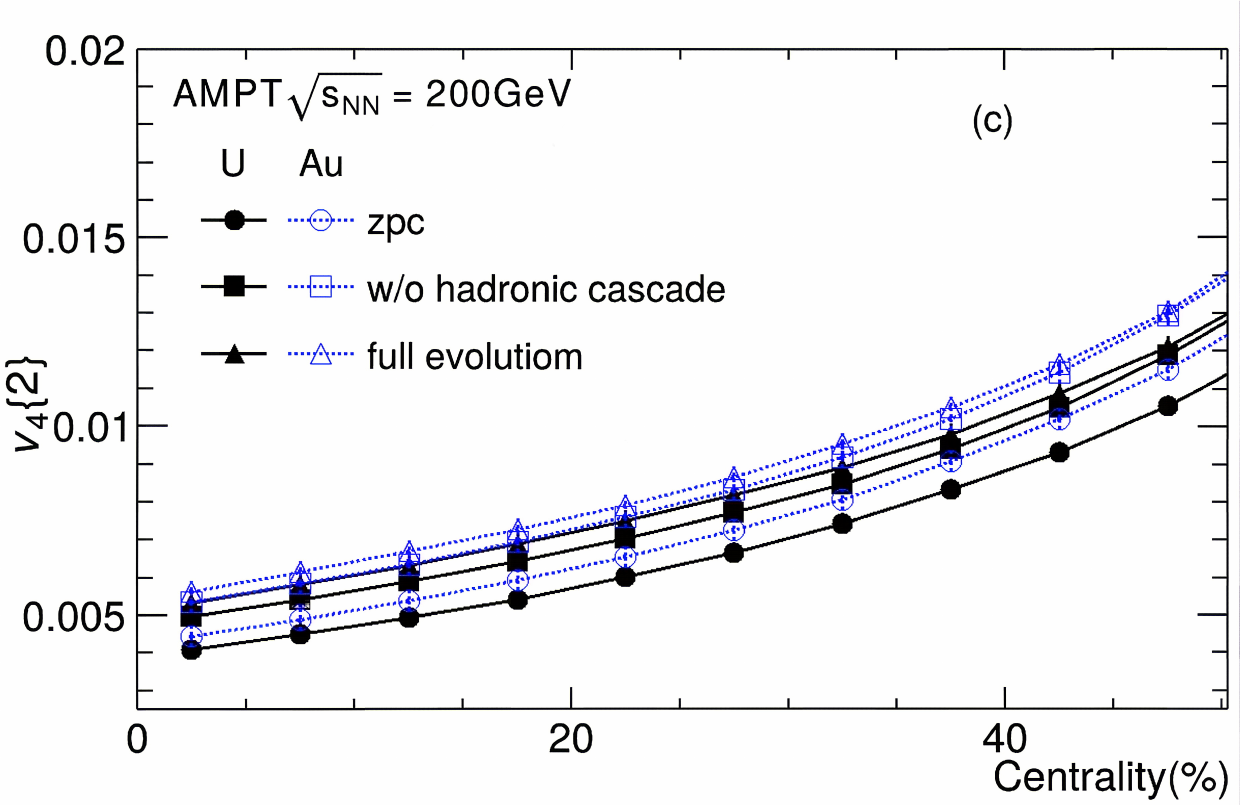}
    \includegraphics[width=0.45\textwidth]{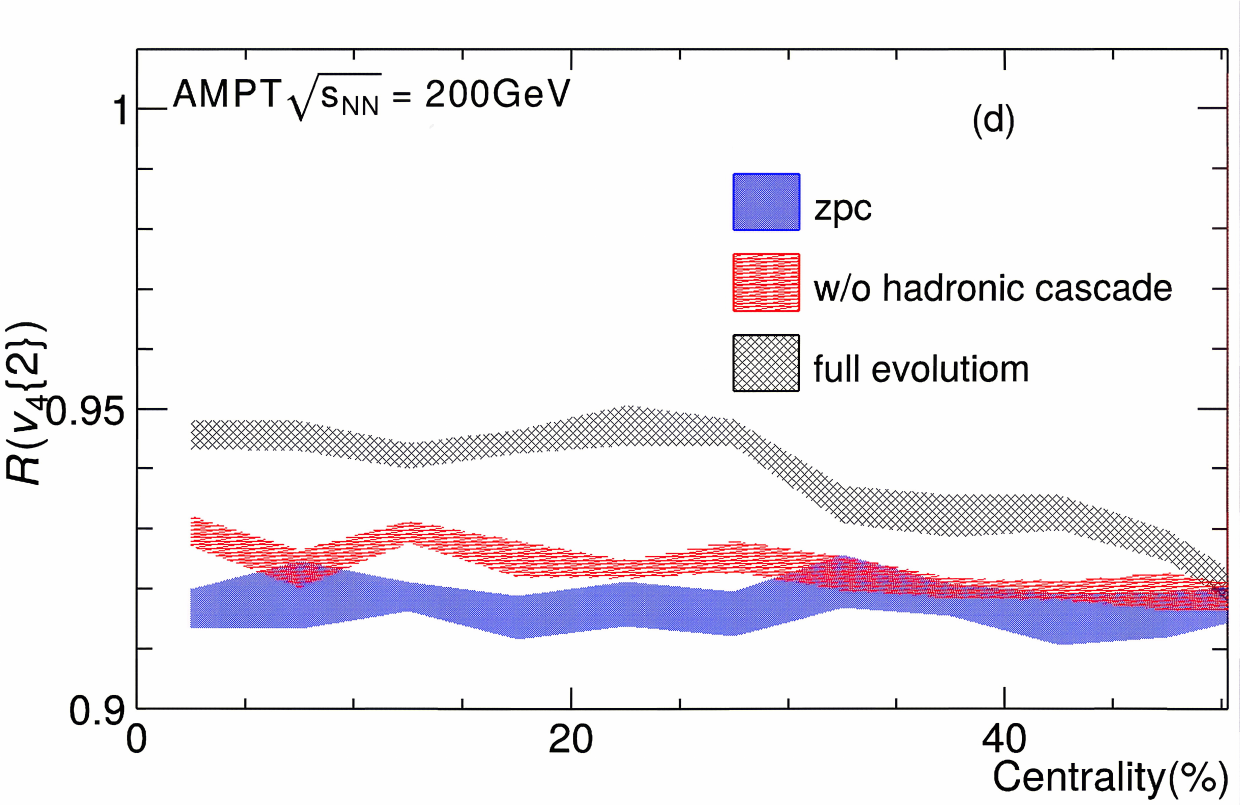}  \\
    \includegraphics[width=0.45\textwidth]{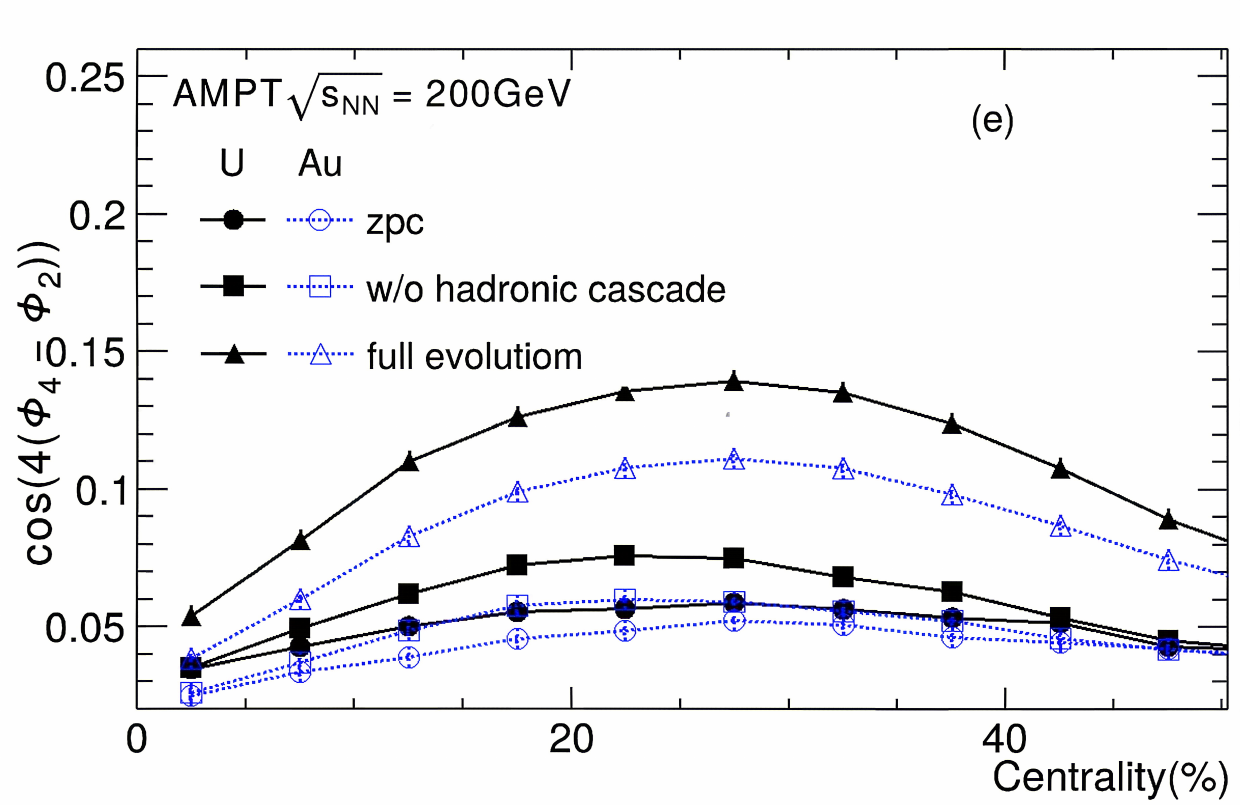}
    \includegraphics[width=0.45\textwidth]{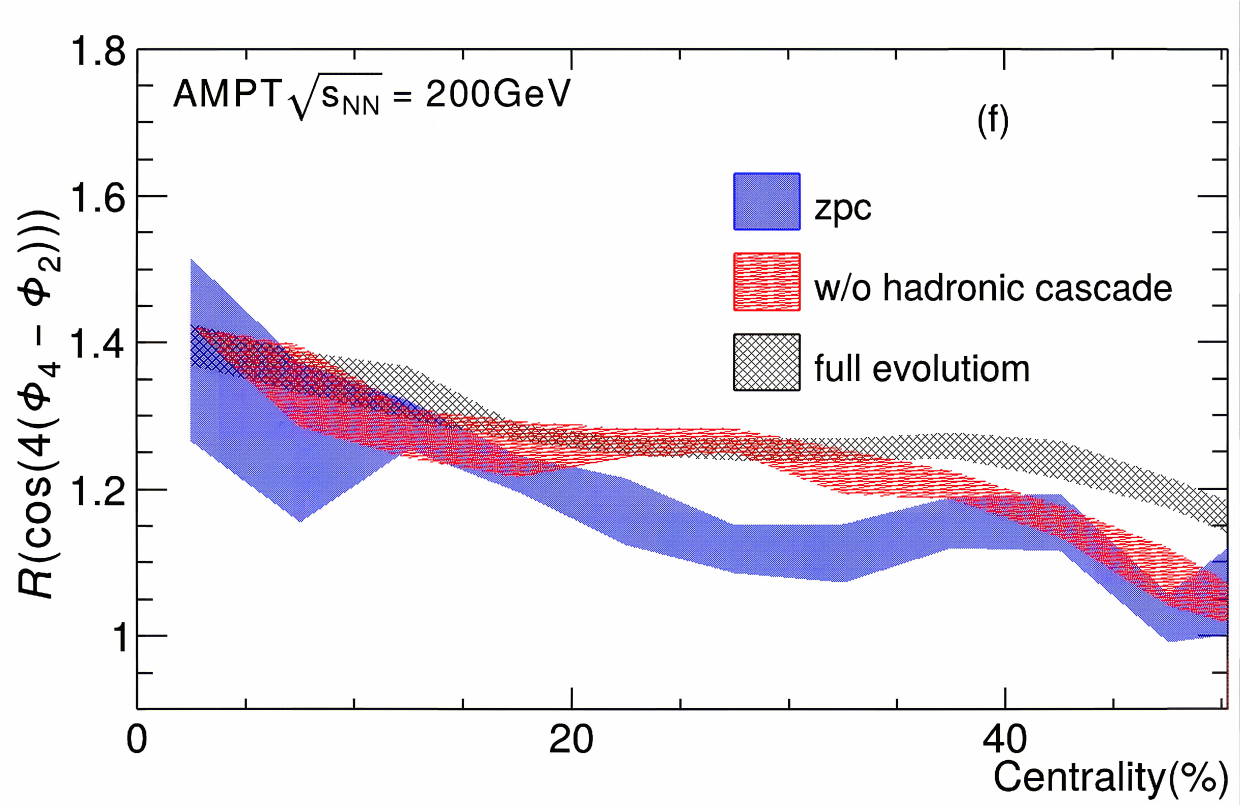}   
    \caption{Similar to Fig.~\ref{fig:chi422}, but for $\vtt$ in panels (a) and (b), $\vft$ in panels (c) and (d), and $\langle \cosDPhi \rangle$ in panels (e) and (f). The shaded bands represent statistical uncertainties.}
    \label{fig:flow}
\end{figure*}

\section{Results and Discussion}

Figure~\ref{fig:chi422}(a) presents the centrality dependence of the nonlinear response coefficient, $\chi_{4,22}$, for U+U and Au+Au collisions separately. To elucidate the microscopic origin and dynamical generation of this observable, the results are extracted at three distinct stages of the AMPT evolution: immediately after the partonic cascade (zpc), after quark coalescence (without hadronic cascade), and after final hadronic rescattering (full evolution). For both systems, the magnitude of $\chi_{4,22}$ exhibits a clear, monotonic increase as the collision system evolves from the early partonic phase to final kinetic freeze-out. This progressive stage-by-stage growth supports the medium-response nature of $\chi_{4,22}$, demonstrating that the nonlinear coupling between elliptic and hexadecapole flow is continuously accumulated through the collective expansion of the medium. Furthermore, in mid-central collisions, $\chi_{4,22}$ shows a relatively weak centrality dependence and behaves similarly in both systems. In the most central collisions, however, $\chi_{4,22}$ in the U+U system shows a visible modification relative to the Au+Au baseline. This behavior is consistent with the sensitivity of $\chi_{4,22}$ to the intrinsic hexadecapole deformation ($\beta_4$) of the uranium nucleus, as suggested by previous studies~\cite{Xu:2024bdh}.

Although the absolute magnitude of $\chi_{4,22}$ is strongly modified by the evolutionary stage of the medium, structural differences between the colliding nuclei can be examined using a relative ratio observable. Figure~\ref{fig:chi422}(b) shows the ratio, $R(\chi_{4,22})$, between the U+U and Au+Au systems over the same centrality range. The underlying physics can be understood by treating the medium as a dynamical amplifier: the absolute magnitude of $\chi_{4,22}$ is determined by the efficiency with which the medium converts initial geometric seeds into nonlinear flow. Although the absolute values of $\chi_{4,22}$ evolve significantly during the collision, taking the relative ratio between two similar systems reduces common medium-response effects, such as the overall amplification efficiency and viscous attenuation. Consequently, $R(\chi_{4,22})$ exhibits approximate stability across the three evolutionary stages over a broad centrality interval. This behavior supports the efficacy of the comparative ratio method as a robust experimental tool for reducing medium-evolution uncertainties and constraining the hexadecapole deformation ($\beta_{\rm 4,U}$) of the colliding nuclei~\cite{Xu:2024bdh,Wang:2024vjf}.

To quantify the stage dependence of $R(\chi_{4,22})$, we perform a constant fit to its values at the three AMPT evolution stages in each centrality bin. The fit results are summarized in Table~\ref{tab:stagefit}. In the centrality range $5$--$45\%$, the obtained $p$-values are generally larger than $0.1$, indicating that the stage dependence of $R(\chi_{4,22})$ is statistically consistent with a constant within the current uncertainties. In contrast, the $0$--$5\%$ and $45$--$50\%$ bins yield small $p$-values, suggesting residual stage dependence in the most central and relatively peripheral collisions. We therefore interpret the stability of $R(\chi_{4,22})$ as an approximate property over a broad centrality range, rather than as exact stage-independent behavior in every bin.

\begin{table}[t]
\centering
\caption{Results of constant fits to $R(\chi_{4,22})$ across the three AMPT evolution stages. The table lists the fitted constant value, $\chi^2/{\rm ndf}$, and the corresponding $p$-value for each centrality bin.}
\label{tab:stagefit}
\begin{tabular}{cccc}
\hline
Centrality & Fit value & $\chi^2/{\rm ndf}$ & $p$ value \\
\hline
$0$--$5\%$   & 0.688 & 5.30 & 0.005 \\
$5$--$10\%$  & 0.842 & 0.31 & 0.733 \\
$10$--$15\%$ & 0.930 & 0.85 & 0.427 \\
$15$--$20\%$ & 0.934 & 0.37 & 0.688 \\
$20$--$25\%$ & 0.961 & 0.59 & 0.557 \\
$25$--$30\%$ & 0.983 & 2.05 & 0.128 \\
$30$--$35\%$ & 0.970 & 1.07 & 0.344 \\
$35$--$40\%$ & 0.980 & 1.39 & 0.248 \\
$40$--$45\%$ & 0.997 & 0.26 & 0.769 \\
$45$--$50\%$ & 0.962 & 8.42 & 0.00022 \\
\hline
\end{tabular}
\end{table}

According to Eq.~(\ref{eq:chi}), the nonlinear response coefficient is mathematically constructed from the three-particle asymmetric cumulant, $\ac$, and the four-particle cumulant of elliptic flow, $\langle v_2^4 \rangle$. To better understand the stability of the $R(\chi_{4,22})$ ratio, it is instructive to examine its components separately, as shown in Fig.~\ref{fig:acandv24}. The general trend indicates that the absolute magnitudes of both $\ac$ and $v_2$ increase as the medium evolves. This behavior is expected: during the collective expansion of the system, the initial spatial geometric anisotropies are progressively converted into the final-state momentum anisotropies of the emitted particles. Interestingly, closer examination shows that although these magnitudes increase over the full duration of the collision, the four-particle cumulant $\langle v_2^4 \rangle$ remains stable during the hadronization process itself. The results extracted immediately after the partonic cascade (zpc) and those extracted immediately after hadronization (without hadronic cascade) nearly overlap. This indicates that the kinematic recombination of partons via coalescence preserves the established elliptic-flow fluctuations.

The behavior of $\chi_{4,22}$ results from a nontrivial interplay between the numerator and the denominator. The quantities $\ac$, $\langle v_2^4 \rangle$, $\vtt$, and $\vft$ all contain substantial flow-magnitude dependence, and their centrality dependences therefore reflect both the evolution of the underlying geometric correlations and the centrality dependence of the absolute flow strength. By contrast, $\chi_{4,22}=\ac/\langle v_2^4 \rangle$ is a normalized nonlinear response coefficient, for which much of the $v_2$-magnitude dependence in the numerator is removed by the denominator. This cancellation explains why $\chi_{4,22}$ exhibits a comparatively flatter centrality dependence in the mid-central region.

The same structure also clarifies the stability of the ratio.
\begin{equation}
R(\chi_{4,22})=
\frac{R({\rm ac}_2\{3\})}{R(\langle v_2^4\rangle)}\,,
\end{equation}
The stability of $R(\chi_{4,22})$ should not be interpreted as evidence for the absence of medium evolution in $\chi_{4,22}$ itself. Rather, both $\ac$ and $\langle v_2^4 \rangle$ are modified during the evolution, but their relative changes between U+U and Au+Au are similar over a broad centrality range. Consequently, the ratio of these two ratios remains comparatively stable. This conclusion is consistent with analytical solutions from relativistic hydrodynamics, which suggest that such nonlinear flow correlations receive contributions from intrinsic event-plane correlations between different orders of initial geometric eccentricities~\cite{Ren:2026fqj}.

In experimental measurements, the nonlinear response coefficient $\chi_{4,22}$ quantifies how strongly the fourth-order flow harmonic is driven by and projected onto the second-order event plane ($\Phi_2$). A broader and more systematic treatment of such nonlinear couplings can be found in the recent study of Ref.~\cite{Ren:2026fqj}. To further disentangle the dynamics of this mode coupling, Fig.~\ref{fig:flow} examines the stage-by-stage evolution of the individual two-particle cumulants, $\vtt$ and $\vft$, together with the explicit flow angular correlation, $\langle \cosDPhi \rangle$. We note that two-particle cumulants and event-plane-based quantities are not strictly identical in the presence of fluctuations and nonflow effects; therefore, they are treated here as related but distinct probes of the flow magnitude and angular correlation. Examining the individual collision systems yields several notable observations regarding the medium evolution. First, during the hadronization process (comparing ZPC with the case without the hadronic cascade), the elliptic flow $\vtt$ remains remarkably stable, whereas the hexadecapole flow $\vft$ exhibits a sizable modification. Conversely, during the final hadronic rescattering phase (full evolution), $\vft$ remains largely unchanged, while $\vtt$ increases substantially.

This distinct and contrasting behavior can be understood from the decomposition $V_4=V_{4L}+\chi_{4,22}V_2^2$. The component $V_{4L}$ represents the direct response to the fourth-order initial eccentricity, whereas the second term is the nonlinear mode-coupled contribution generated from elliptic flow. The nonlinear component is directly constrained by the multiparticle correlations entering $\chi_{4,22}$. In contrast, the linear component $V_{4L}$ is usually obtained indirectly by combining the measured $v_4$ with the nonlinear contribution, which introduces additional assumptions, as discussed in Ref.~\cite{Ren:2026fqj}. A detailed stage-by-stage separation of $V_{4L}$ and the nonlinear contribution is therefore left for future work.

Despite these varied responses in the individual flow harmonic magnitudes, the event-plane correlation itself, $\langle \cosDPhi \rangle$, is found to increase monotonically through both the hadronization and hadronic rescattering stages. This dynamical enhancement of the angular correlation is consistent with the continuous growth of $\chi_{4,22}$ observed in Fig.~\ref{fig:chi422}, indicating that the correlation between the second- and fourth-order momentum anisotropies is progressively built up by the medium expansion. Turning to the relative ratios between the U+U and Au+Au systems, we find that $R(\chi_{4,22})$ is more stable than the ratios involving $\vft$ and $\langle \cosDPhi \rangle$, which retain visible residual stage dependence. This distinction shows that the approximate stage independence is a specific feature of the normalized nonlinear response ratio rather than a generic property of all related observables.

To estimate the possible influence of nonflow and self-correlation effects, we also checked the analysis using a subevent method. The resulting $R(\chi_{4,22})$ is consistent with the full-event result within statistical uncertainties over most of the centrality range, indicating that the observed stability of $R(\chi_{4,22})$ is not primarily caused by nonflow effects associated with the full-event construction. Residual nonflow contributions cannot be completely excluded, especially in the most central and relatively peripheral bins, and should be further controlled in future precision analyses~\cite{Borghini:2000cm, Wang:2008gp}.

\section{Summary}

In summary, we have presented a systematic investigation of the nonlinear response coefficient, $\chi_{4,22}$, in U+U and Au+Au collisions at $\snn = 200$ GeV using the AMPT model. By extracting observables at distinct snapshots of the collision evolution---specifically after the partonic cascade, immediately after quark coalescence, and after final hadronic rescattering---we explicitly tracked the dynamical generation of nonlinear mode coupling. Our results demonstrate that the absolute magnitudes of $\chi_{4,22}$, its constituent components ($\ac$ and $\langle v_2^4\rangle$), and the flow angular correlation $\langle \cosDPhi \rangle$ increase as the medium evolves. This supports the interpretation that the nonlinear coupling between elliptic and hexadecapole flow is a characteristic medium response, progressively built up through collective partonic and hadronic expansion.

Despite the strong dependence of these absolute magnitudes on the evolutionary stage of the medium, the ratio $R(\chi_{4,22})$ between the U+U and Au+Au systems exhibits approximate stage independence over a broad centrality range. A constant-fit test shows that this stability is statistically supported mainly in the $5$--$45\%$ centrality interval, while residual deviations remain in the $0$--$5\%$ and $45$--$50\%$ bins. Detailed differential analyses reveal that, although individual flow harmonics are subject to complex modifications from hadronization kinematics and hadronic dissipation, the normalized ratio $R(\chi_{4,22})$ reduces common medium-response effects, such as overall flow amplification and viscous attenuation, more effectively than several related observables. The comparison between the full-event and subevent methods further indicates that the observed stability is not dominated by nonflow effects, although residual nonflow cannot be fully excluded.

These findings support the use of $R(\chi_{4,22})$ as a nuclear-structure-sensitive observable that can help constrain the intrinsic hexadecapole deformation ($\beta_4$) of atomic nuclei, in a manner consistent with previous deformation studies. Since the present work does not perform an independent controlled scan over $\beta_4$, the result should be regarded as a stage-by-stage AMPT robustness check of the ratio observable rather than a standalone extraction of $\beta_4$. Moving forward, applying this comparative methodology to high-statistics experimental data, together with dedicated deformation scans and improved nonflow suppression, will provide a more precise resolution of the higher-order spatial structure of deformed nuclei.

\section*{Acknowledgments}
We thank Fuqiang Wang for fruitful discussions.
This work was supported in part by the National Natural Science Foundation of China under Grants No.~12275082, No.~12275053, No.~12035006, No.~12205310, and No.~12147101.

\bibliography{myref}

\end{document}